\documentclass[twocolumn]{webofc}

\usepackage[varg]{txfonts}  
\usepackage{hyperref}
\usepackage{url}
\let\clearpage\relax
\hypersetup{colorlinks=true,citecolor=blue,urlcolor=blue,linkcolor=blue}
\begin{document}
\title{Lead tungstate calorimeter of the  Jefferson Lab Eta Factory experiment}

\author{\firstname{Alexander} \lastname{Somov}\inst{1}\fnsep\thanks{\email{somov@jlab.org}}\fnsep\thanks{For the GlueX collaboration and JEF working group}
}

\institute{Thomas Jefferson National Accelerator Facility, Newport News, Virginia, 23606, USA
          }

\abstract{
A new electromagnetic calorimeter (ECAL) consisting of 1596 lead tungstate PbWO$_{\rm 4}$  scintillating crystals has 
been fabricated and installed in the experimental Hall D at Jefferson Lab (JLab). The high-granularity, high-resolution 
calorimeter is required by the JLab Eta Factory experiment, whose main physics goal is to study rare decays of eta mesons. 
The ECAL replaced the inner part of the forward lead glass calorimeter of the GlueX detector. Signals from the detector 
will be digitized using twelve-bit flash analog-to-digital converters operated at a sampling rate of 250 MHz. The ECAL is 
integrated into the trigger system of the GlueX detector using electronics modules designed at JLab. The ECAL is currently 
at the commissioning stage and should be ready for the physics run in January 2025. We will give an overview of the JEF 
experiment, the design and construction of the ECAL, and the integration of the detector and its infrastructure into the 
Hall D experimental setup. 
}

\maketitle

\section{Introduction}
\label{intro}
The GlueX detector~\cite{gluex} in experimental Hall D at Jefferson Lab (JLab) was designed to perform experiments using a 
photon beam incident on various targets. The main goal of the JLab Eta Factory (JEF) experiment~\cite{jef,jef1} is to perform 
measurements of various $\eta^{(\prime)}$ decays with emphasis on rare neutral modes. The physics program includes searches for gauge 
bosons coupling the dark sector to the Standard Model sector, precision tests of low-energy QCD, determination of the quark mass 
ratio, and others. The experiment requires good reconstruction of photons in the forward direction, which can be achieved by 
using an electromagnetic calorimeter (ECAL) based on high-granularity high-resolution lead tungstate  PbWO$_{\rm 4}$  scintillating 
crystals. The ECAL replaced the inner part of the GlueX forward lead glass calorimeter. Prior to the construction of the large-scale 
ECAL detector, we built a prototype, which was used to study performance of ECAL modules and light monitoring system, and to optimize 
the design of the front-end electronics for the JEF operating conditions. The prototype consisted of an array of 12 $\times$ 12 
PbWO$_{\rm 4}$ modules and was successfully used in the PrimEx-$\eta$ experiment in Hall D to reconstruct Compton scattering events. 
The performance of the prototype is presented in Ref.~\cite{ccal}. In this paper we will describe the design, fabrication, and installation 
of the ECAL on the forward GlueX calorimeter and also give an overview of the ECAL electronics and integration of the detector into the 
GlueX trigger system.

The construction of the ECAL started in 2022 and the detector will be fully commissioned in the early fall of the 2024. The total cost of 
the project is about 5 million dollars.

\section{Calorimeter design}
\begin{figure}[]
\centering
\includegraphics[width=0.95\linewidth]{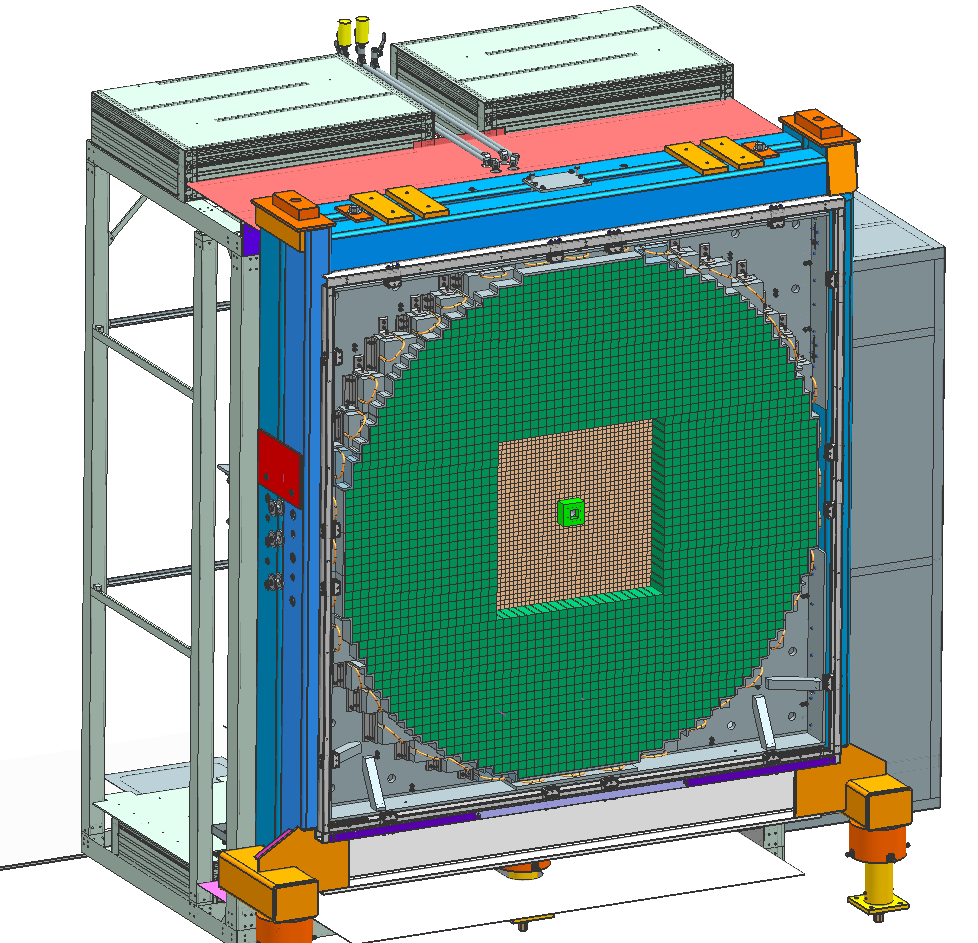}
\caption{The frame of the GlueX forward calorimeter with the ECAL modules (brown rectangular area in the center) surrounded by lead glass modules (green). 
The photon beam passes through the hole in the middle of the ECAL. The innermost layer of the ECAL is covered by an absorber, shown as a light green area 
in the center of the ECAL.}
\label{fig:design}
\end{figure}
\begin{figure*}[]
\centering
$\vcenter{\hbox{\includegraphics[width=0.33\linewidth]{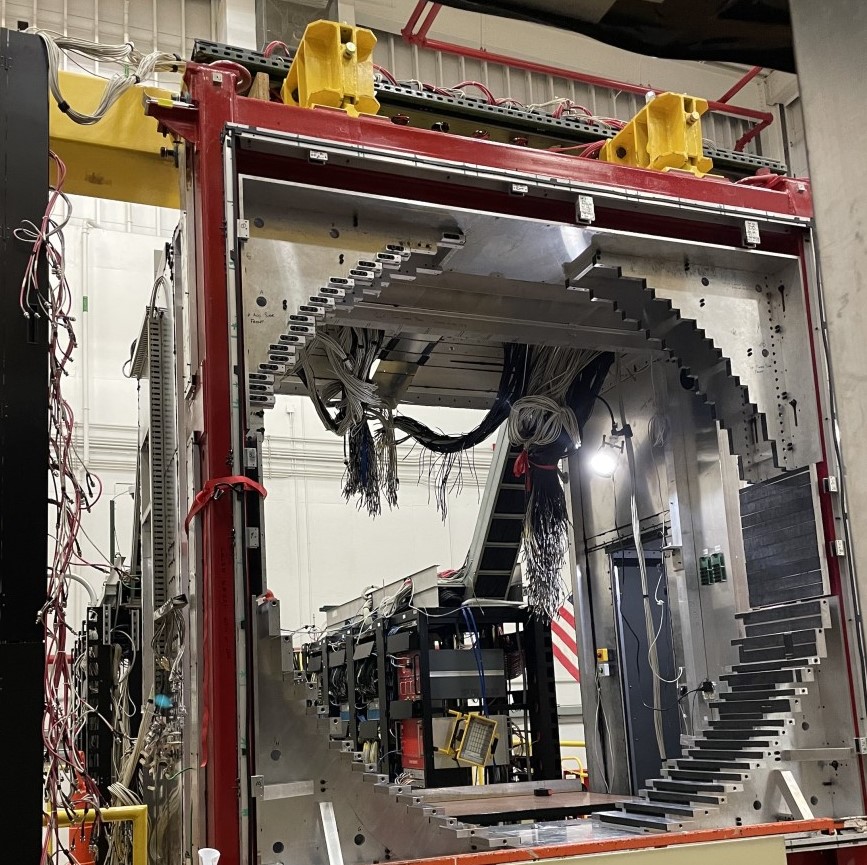}}}$
\hspace{0.05in}
$\vcenter{\hbox{\includegraphics[width=0.3\linewidth]{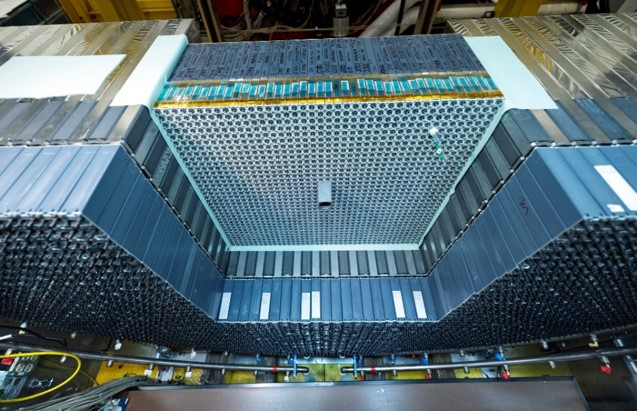}}}$
\hspace{0.05in}
$\vcenter{\hbox{\includegraphics[width=0.33\linewidth]{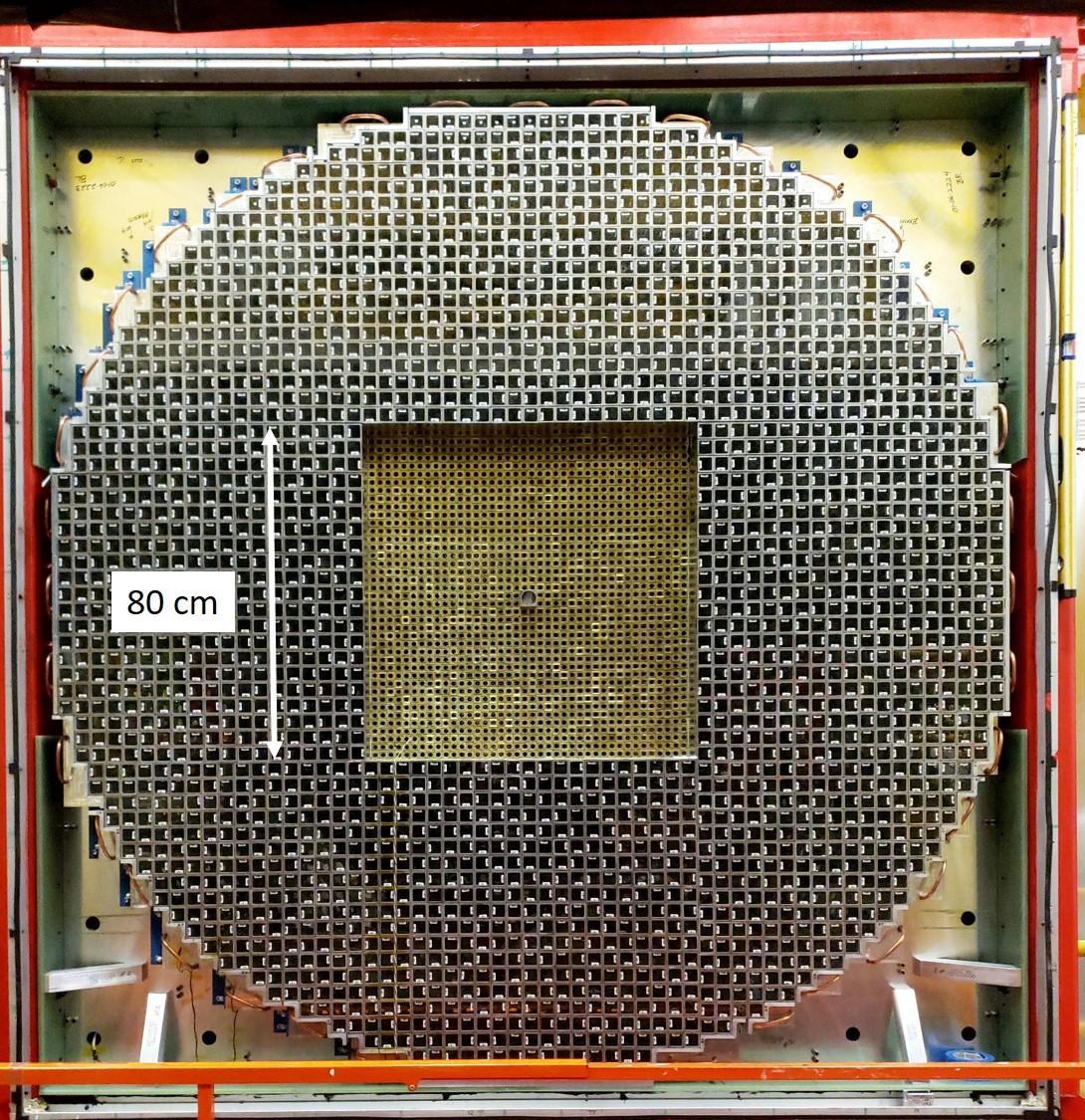}}}$
\caption{Installation of the ECAL modules on the GlueX forward calorimeter. The original calorimeter frame after removing all lead glass modules (left). 
Installation of the ECAL surrounded by the lead glass modules (middle). Installed calorimeter with the ECAL in the center of the detector (right).}
\label{fig:instal}
\end{figure*}
The ECAL consists of 1596 lead tungstate rectangular scintillating crystals. Each crystal has the following dimensions: 
2.05 cm $\times$ 2.05 cm $\times$ 20 cm. The calorimeter modules are organized into an array of 40 $\times$ 40 modules with a hole of 
2 $\times$ 2 modules in the middle for the beam. The PbWO${\rm_4}$ Type II crystals were purchased from two vendors:  SICCAS (China) and 
CRYTUR (the Czech Republic). The properties of crystals used for the ECAL fabrication were studied in Ref.~\cite{crystals}. The main characteristics 
of the CRYTUR crystals, such as radiation resistance, light yield, and optical transmission, appeared to be somewhat more uniform compared to 
those from SICCAS. Due to the good radiation properties, we installed 918 available CRYTUR crystals around the beamline, where 
the detector will be exposed to higher radiation levels.

The ECAL replaced the inner part of the GlueX forward lead glass calorimeter, which originally consisted of 2800 modules with a block size of 4 cm 
$\times$ 4 cm $\times$ 45 cm. The schematic view of the hybrid PbWO$_4$ and lead glass calorimeter is presented in Fig.~\ref{fig:design}. In this plot, 
the ECAL is the rectangular brown area surrounded by the lead glass modules shown in green. The photon beam passes through the hole in the middle of 
the ECAL. The innermost ECAL layer around the beam pipe is shielded by a 6 cm thick tungsten absorber. All modules are placed inside the detector frame. 
The calorimeter is positioned about 6 m downstream from the target and covers a polar angle between $0.2^\circ$ and $11^\circ$ for photons originating 
from the target. The smaller transverse size of the PbWO${\rm_4}$ crystals relative to the lead glass blocks will improve the separation of showers in 
the forward direction. The typical energy resolution of the lead glass calorimeter is $\sigma_{\rm E}/E = 6.2\%/\sqrt E \oplus 4.7\%$~\cite{gluex}, which 
is more than a factor of two worse compared to the ECAL. The radiation resistance of lead tungstate crystals is superior to lead glass, which is important 
for the long-term operation of the calorimeter at high luminosity. 

During the installation of the ECAL, the lead glass calorimeter had to be fully disassembled. Since the light output of the  PbWO${\rm_4}$ crystals depends 
on temperature, the frame surrounding the detector modules was modified by installing cooling blocks with  build-in pipes to circulate cooling liquid. 
The size and position of the cooling blocks in the frame were chosen to accommodate the different size of the PbWO$_4$ and lead glass blocks. The detector 
frame was designed and installed by the Hall D technical and engineering group. The entire detector was positioned  inside a foam-insulated, 
light-tight box. During operation, the detector will be maintained at a temperature of approximately  17$^\circ$ C with expected stability better 
than 0.2$^\circ$ C. 

After installing the new detector frame, lead glass and ECAL modules were stacked together. The module installation was organized into several groups of 
layers. After installing each group of layers, the modules were squeezed by a movable cooling block from one side of the detector. A thermally conductive 
silicon pad (Saint-Gobain ThermalCool) with a compressed thickness of about 600 $\mu$m was applied as a compliance layer between the lead tungstate insert 
and lead glass modules. The main steps of the detector installation are presented in Fig.~\ref{fig:instal}, with the fully installed calorimeter shown on 
the right plot.
\begin{figure*}[]
\centering
\includegraphics[width=0.47\linewidth]{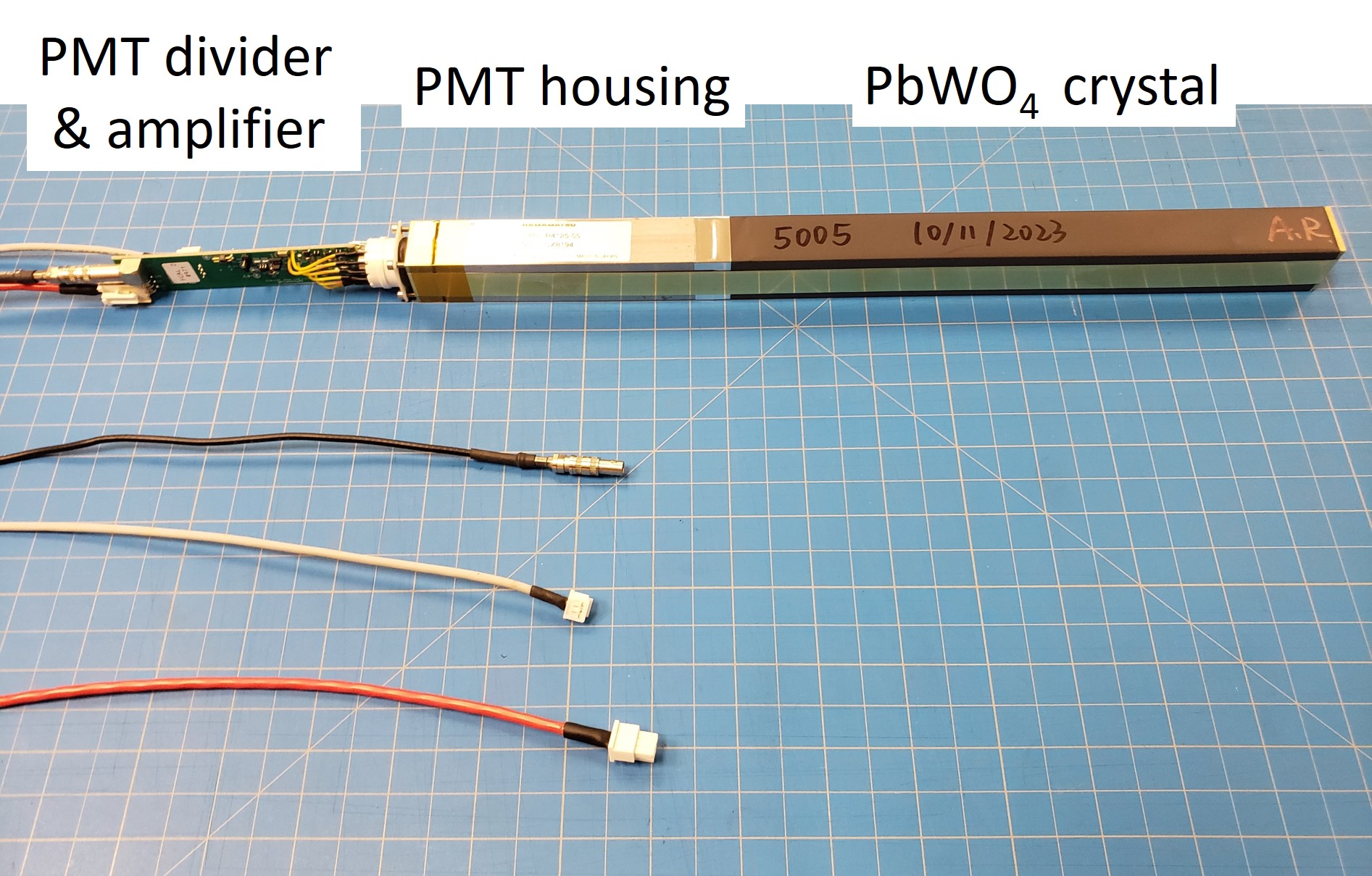}
\includegraphics[width=0.52\linewidth]{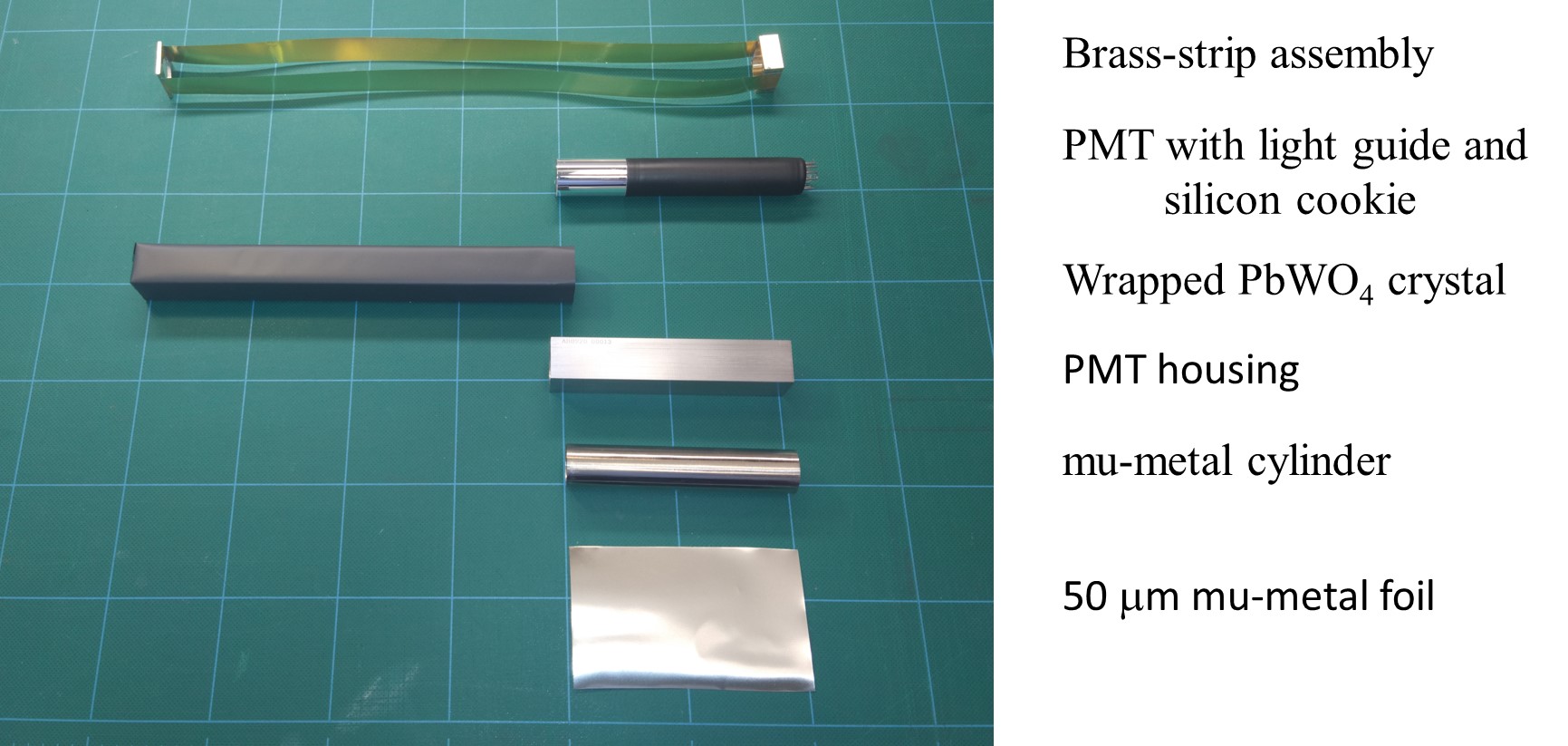}
\caption{Assembled ECAL module with the attached PMT active base with signal, high voltage, and low voltage cables (left plot). Main components of the ECAL 
module (right plot):  the PbWO$_{\rm 4}$ crystal wrapped with the ESR reflective foil and Tedlar, PMT with the light guide, PMT housing, brass-strip 
assembly, and mu-metal shielding.}
\label{fig:module}
\end{figure*}
\section{Module design}
The ECAL has a modular structure. An assembled calorimeter module and its main components are presented in Fig.~\ref{fig:module}. A lead tungstate crystal 
wrapped with a 65 $\mu$m thick reflective ESR foil and light-tight Tedlar is connected to a 3.5 cm long acrylic light guide using a 1.8 mm thick silicon 
rubber disk (cookie). The other end of the light guide is glued to a Hamamatsu 4125 photomultiplier tube (PMT) using Dymax 3094 UV curing glue. The PMT is 
surrounded by a 350 $\mu \rm m $-thick Amumetal cylinder and a  50 $\mu \rm m $-thick mu-metal foil and is positioned inside a soft iron housing. 
The cylinder is separated from the PMT housing using a thin Kapton film.

The PMT housing is required to reduce the fringe field of the solenoid magnet present in the calorimeter region, while the light guide extends 
the shielding above the face of the PMT photo cathode. The maximum magnetic field components directed along the PMT (longitudinal) and perpendicular to the 
PMT (transverse) are about 60 Gauss and 10 Gauss, respectively.  The PMT magnetic shielding was studied using the TOSCA magnetic field simulation program 
and verified with detector prototypes positioned inside magnetic field produced by Helmholtz coils. The performance of the PMT in the magnetic field was 
studied using an LED pulser. 

The module components are held together using a brass-strip assembly, which consists of two 50 $\mu \rm m $ thick brass strips soldered to two brass 
flanges. One flange is placed on the crystal face, and the other on the PMT housing. Four set screws in the PMT housing flange stretch the strips 
and hold the components together. A PMT with the light guide is pushed towards the silicon cookie by a G-10 retaining plate with four screws connected 
to the housing flange.

\section{Electronics and trigger system}
\begin{figure*}[]
\centering
$\vcenter{\hbox{\includegraphics[width=0.48\linewidth]{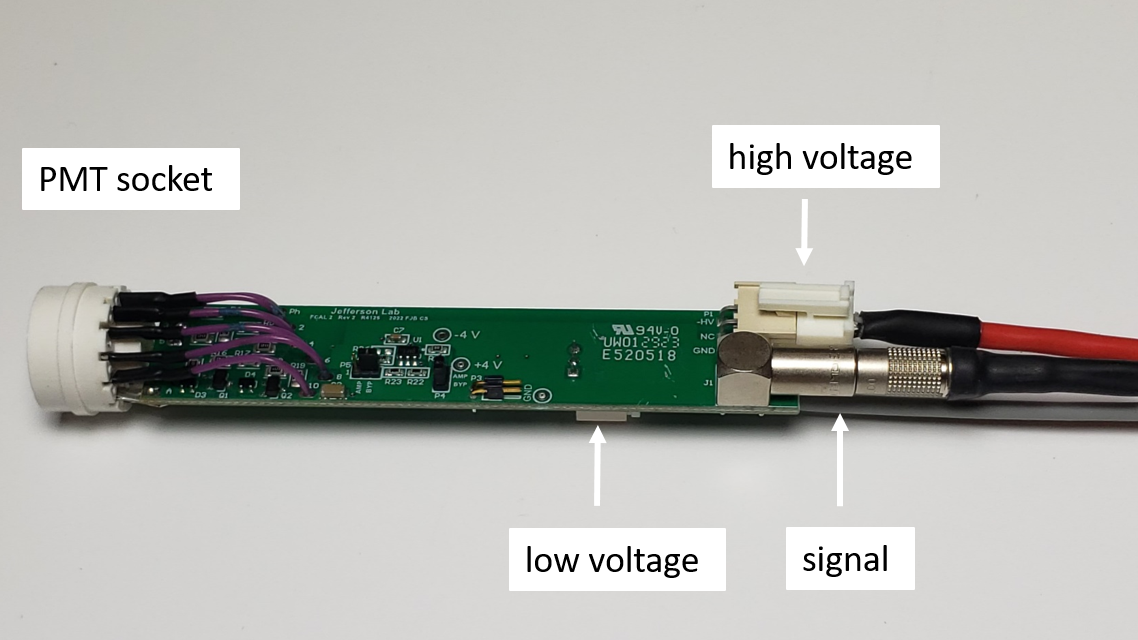}}}$
\hspace{0.05in}
$\vcenter{\hbox{\includegraphics[width=0.48\linewidth]{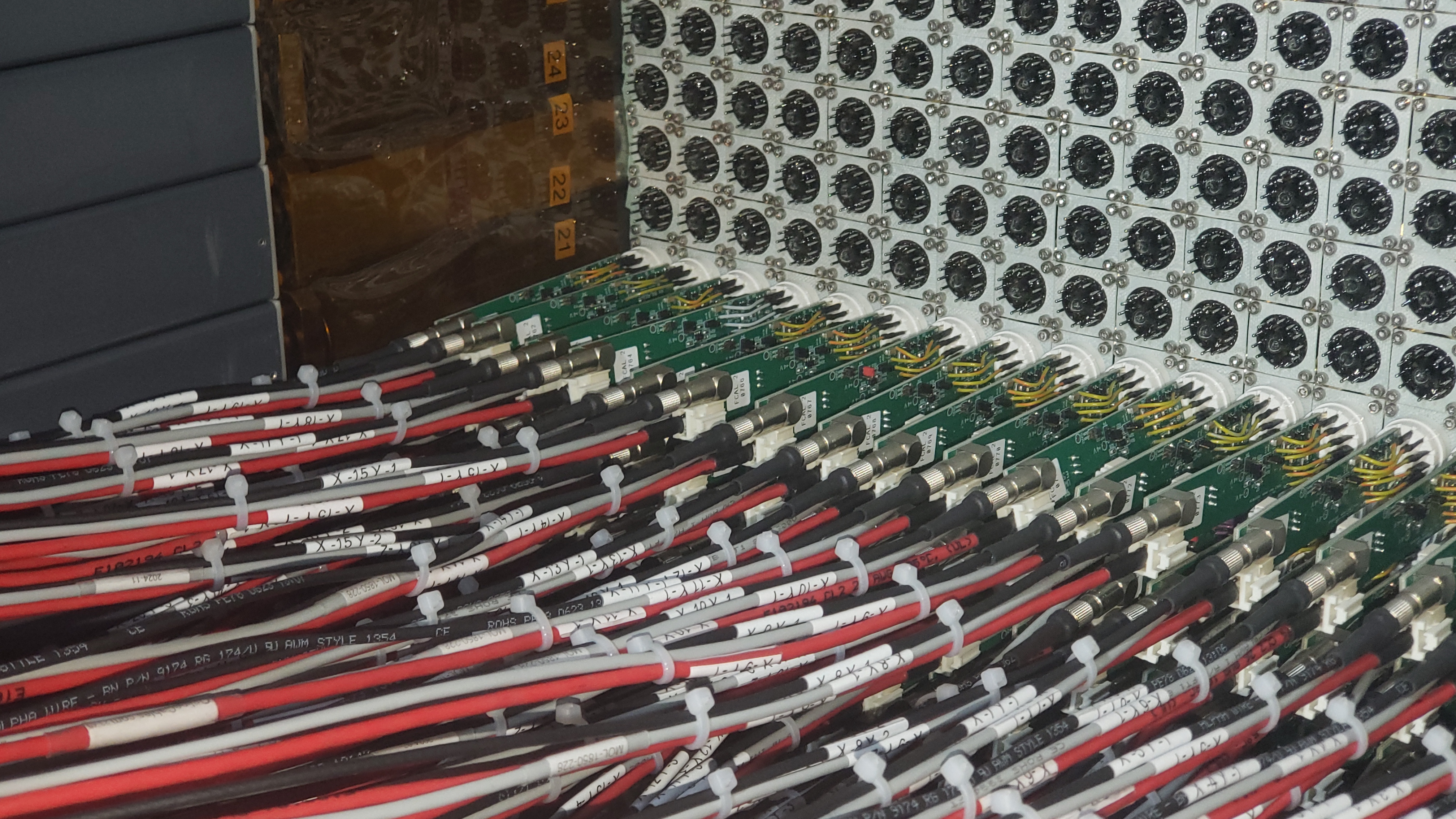}}}$
\caption{PMT active bases designed at Jefferson lab (left). Installation of the active bases on the ECAL PMTs (right). Each base is 
connected to signal, high voltage, and low voltage cables.}
\label{fig:active_base}
\end{figure*}

The distribution of high voltages (HV) to the PMT dynodes and the readout of electrical pulses are managed by the PMT active base designed at Jefferson Lab. 
The base incorporates a PMT divider and an amplifier positioned on the same printed circuit board (PCB). To improve gain stabilization at high rates, the 
original Hamamatsu divider design for this type of PMT was modified by adding two bipolar transistors to the last two dynodes. The amplifier, with a 
gain of about three, allows for a lower PMT operating voltage, thereby reducing the anode current, which  also improves the rate capability and prolongs 
the PMT’s life. This is particularly important for the ECAL modules positioned close to the beamline, which are exposed to the large rate of several 
hundred kilohertz at the readout threshold. The operational-based amplifier requires $\pm$ 5 V power, which is supplied to the active bases by two 
WIENER MPOD modules. The typical high voltage applied to the PMT active base is about 1 kV. The high voltage for each PMT  is provided by a 48-channel 
CAEN A7030N module, with thirty six modules positioned in three SY4527 mainframes. The active base with the attached signal, high voltage, and low 
voltage cables is presented in the left plot of Fig.~\ref{fig:active_base}. Before installing on the detector, all fabricated ECAL modules and dividers 
were tested using an LED test setup. Installation of the active bases on the ECAL PMTs is presented in the right plot of Fig.~\ref{fig:active_base}.

Signals from the PMT are digitized using a twelve-bit 16-channel flash ADC designed at JLab. The flash ADC operates at a sampling rate of 250 MHz 
and is equipped with Field-Programmable Gate Array (FPGA) chips, which allow programming the readout and trigger processing algorithms. 
One hundred flash ADCs are positioned in 7 VXS crates. The VXS (ANSI/VITA 41.0) crate extends the VME/VME64x architecture by implementing 
a high-speed bus used to transmit flash ADC data to the trigger module located in the middle of the crate. The digitized amplitudes (energies) in each 
flash ADC are stored in the FPGA-based pipeline for the VME readout and are summed over all  16 channels and sent to the VXS trigger processor module 
(VTP). The VTP processes amplitudes received from all flash ADCs in the crate and sends the resulting sum via a fiber optical link to the global trigger 
processor situated in the trigger crate, which makes the final trigger decision. The main physics trigger of the JEF experiment is based on the energy 
depositions in the forward and barrel calorimeters. The core module of the trigger system is the trigger supervisor (TS). The TS sends triggers to all 
readout crates to initiate the data readout, provides 250 MHz clock and synchronization signals.  All information from the TS is distributed to GlueX 
crates through optical links. The GlueX trigger system has a fixed latency of about 3 $\mu$s, this is the time needed to make the trigger decision and 
distribute it to the readout modules. The typical trigger rate of the GlueX experiments is 60-70 kHz, with a data rate from the detector of about 1.3 Gbps.
\begin{figure}[hbt]
\centering
\includegraphics[width=0.68\linewidth]{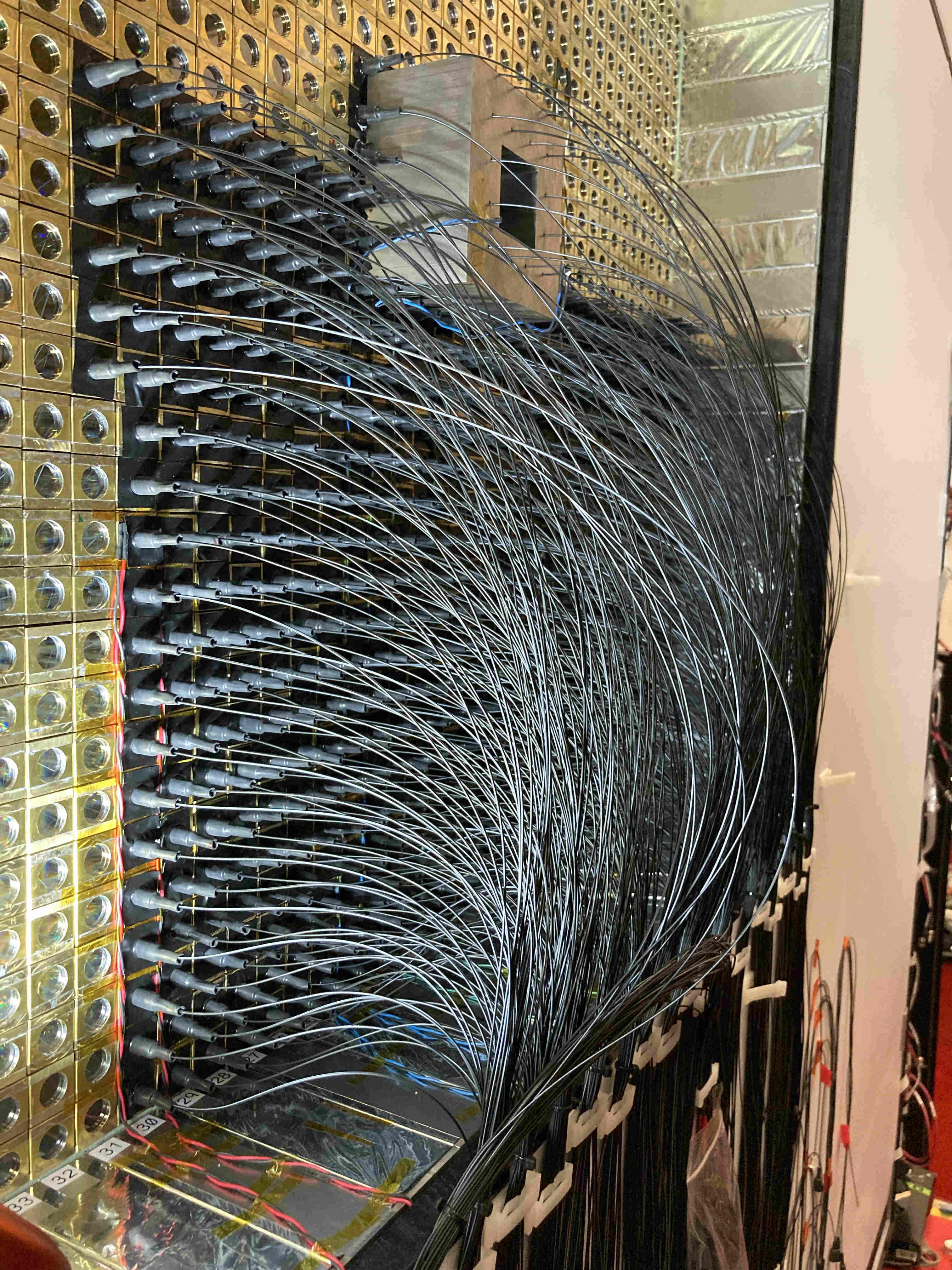}
\caption{Installation of the LMS optical fibers. Optical fiber is glued to each PbWO$_4$ crystal through the hole in the flange on the face of 
the crystal.}
\label{fig:lms}
\end{figure}
\section{Light monitoring system}
The performance of each calorimeter module during data taking will be monitored using an LED-based Light Monitoring System (LMS). 
Light induced by multiple blue LEDs is mixed in an Integrating Sphere (Edmund Optics) with a diameter of 6 inches and distributed to 
each crystal using a plastic optical fiber with a core diameter of 500 $\mu \rm m$. The surface-mount LEDs are soldered onto a PCB, which 
is attached to the input optical port of the integrating sphere. All optical fibers are bundled and connected to the output port of the sphere. 
On the detector side, each fiber is attached to a small plastic cap, which is glued to the face of  a PbWO$_4$ crystal using Dymax 3094 UV 
curing glue. After gluing, the face of each module is closed by light-tight tape to prevent optical crosstalk between modules. The installation 
of the optical fibers to ECAL modules is shown in Fig.~\ref{fig:lms}. The stability of the LEDs is monitored using two reference PMTs that 
receive light from two sources: (1) a single fiber from LEDs and (2) a YAP:Ce scintillating crystal glued to the PMT activated by $^{241}$Am 
$\alpha$-source. The LEDs and reference PMTs are positioned inside the calorimeter temperature-stabilized room. The LED driver allows control over 
the LED pulse amplitudes and rates. The LMS is integrated into the GlueX trigger system and will be continuously operated with a typical rate of 
10 Hz during data taking.
\section{Summary}
A new electromagnetic calorimeter based on 1596 PbWO$_4$ scintillating crystals has been fabricated and installed in experimental Hall D 
at Jefferson Lab. The ECAL will significantly improve the reconstruction of photons in the forward direction, thereby extending the physics 
potential of the GlueX detector. The ECAL is currently undergoing commissioning and will be ready for the first physics run in January 2025.

\section{Acknowledgments}
This work was supported by the Department of Energy, USA. Jefferson Science Associates, LLC operated Thomas Jefferson National 
Accelerator Facility for the United States Department of Energy under contract DE-AC05-06OR23177. We are thankful to the JLab's physics division groups 
and members from participating universities for their assistance with the ECAL fabrication and installation of the detector in the experimental hall.

\end{document}